\documentclass[a4paper,fleqn]{cas-dc}

\usepackage{graphicx} 
\usepackage{amsmath} 
\usepackage{amsthm}
\usepackage{braket}

\usepackage{cite} 
\usepackage{siunitx} 

\def\tsc#1{\csdef{#1}{\textsc{\lowercase{#1}}\xspace}}
\tsc{WGM}
\tsc{QE}

\begin{document}
\let\WriteBookmarks\relax
\def\floatpagepagefraction{1}
\def\textpagefraction{.001}

\shorttitle{Integrated Deflector Shield Technology for Spacecraft}    

\shortauthors{Florian Neukart}  

\title [mode = title]{Integrated Deflector Shield Technology for Spacecraft}  



%
\author[1]{Florian Neukart}[type=author,
      style=chinese,
      auid=000,
      bioid=1,
      orcid=0000-0002-2562-1618]


\affiliation[1]{organization={Leiden Institute of Advanced Computer Science},
            addressline={Snellius Gebouw, Niels Bohrweg 1}, 
            city={Leiden},
            postcode={2333 CA}, 
            state={South Holland},
            country={Netherlands}}


\begin{abstract}
The increasing velocity and frequency of space missions necessitate advancements in spacecraft protection technologies to ensure mission success and the safety of onboard systems and personnel. Existing shielding solutions, including electromagnetic shields, plasma shields, and speculative force fields, each offer distinct protective capabilities but are limited when used independently. This study proposes an integrated deflector shield system that combines the strengths of these technologies to provide comprehensive protection against a wide range of spaceborne threats, including charged particles, micrometeoroids, and high-energy radiation. By leveraging fusion energy as a primary power source, we aim to sustain a high-energy shield capable of dynamically adjusting to varying threat levels. Extensive theoretical modeling, simulations, and analytical calculations demonstrate the feasibility of this integrated approach, highlighting significant improvements in energy efficiency and scalability. The implementation of this multi-layered shield system marks a pivotal advancement in spacecraft protection, paving the way for safer and more reliable space exploration missions.
\end{abstract}

\begin{highlights}
\item Proposes an integrated deflector shield system for spacecraft combining electromagnetic, plasma, and speculative force field technologies.
\item Develops mathematical models to evaluate the interactions and combined effectiveness of the integrated shield components.
\item Highlights the potential of leveraging fusion energy to sustain the high energy requirements of the shield system.
\item Identifies key challenges, including energy consumption, material durability, and technological feasibility of the force field component.
\item Provides theoretical analysis on the deflection of charged particles, energy absorption from micrometeoroids, and the speculative repulsive forces from force fields.
\item Suggests future research directions focusing on advanced simulations, experimental validation, material science advancements, and fusion energy integration.
\item Explores potential interdisciplinary applications and benefits of the integrated shield technology beyond space missions.
\end{highlights}


\maketitle

\sloppy

\section{Introduction}
Space exploration presents numerous challenges, one of the most significant being the protection of spacecraft from the harsh environment of space. Traditional shielding technologies have provided partial solutions, but as missions become more ambitious and spacecraft travel at higher speeds, there is a growing need for more advanced protective measures. This section provides an overview of existing shielding technologies, their limitations, and the objectives of our study to develop an integrated deflector shield system.

\subsection{Background}
The exploration and utilization of space have significantly progressed over the past decades, necessitating advanced protection systems for spacecraft. Traditional shielding technologies, such as electromagnetic and plasma shields, have been developed to protect spacecraft from charged particles, micrometeoroids, and cosmic radiation. Electromagnetic shields utilize magnetic fields to deflect charged particles, offering a reliable means of protection against solar and cosmic radiation \cite{Vogt2003}. However, their effectiveness is limited against neutral particles and high-velocity debris.

Plasma shields, which involve the generation of a plasma layer around the spacecraft, aim to absorb and dissipate energy from micrometeoroid impacts. Early research has shown potential in using plasma sheaths for impact mitigation \cite{Krall1973,Fortov2005}. Nonetheless, plasma shield technology remains in its infancy, facing challenges in stability and energy efficiency \cite{Hastings1995}.

Force fields, a concept popularized by science fiction, remain largely speculative. While theoretical frameworks exist, practical implementation is constrained by current technological limitations \cite{Everett2005}.

Despite the progress made in individual shielding technologies, the need for a comprehensive protection system that integrates multiple methods has become increasingly apparent. The limitations of each technology highlight the potential benefits of an integrated approach, combining the strengths of electromagnetic, plasma, and theoretical force fields to provide robust and adaptive protection for spacecraft.

Recent advancements in propulsion technologies, such as the Magnetic Fusion Plasma Drive \cite{Neukart2024}, also underscore the need for more sophisticated shielding systems to support long-duration and high-speed missions. Additionally, efforts to colonize Mars and other celestial bodies further emphasize the importance of developing reliable protection systems for spacecraft and habitats in harsh space environments \cite{Neukart2024b}.

\subsection{Objectives}
This study aims to develop an integrated deflector shield system that combines electromagnetic, plasma, and theoretical force field technologies to provide comprehensive spacecraft protection. The proposed system can dynamically adjust shield parameters to counter varying threats by leveraging the high-energy output of fusion drives, for example. The key objectives of this research are:

\begin{enumerate}
    \item To develop mathematical models that describe the interaction between different shield types.
    \item To conduct extensive theoretical analysis to evaluate the integrated system's performance.
    \item To propose design frameworks that validate the theoretical models.
    \item To analyze the energy efficiency and scalability of the proposed shield system.
\end{enumerate}

The successful implementation of this integrated deflector shield technology will significantly enhance the safety and reliability of high-speed spacecraft, facilitating the advancement of space exploration missions.

\section{Literature Review}
The development of effective deflector shield technology for spacecraft has been the focus of extensive research over the years. Various approaches, including electromagnetic shields, plasma shields, and speculative force fields, have been proposed and explored to address the diverse threats faced in space. This section reviews the current state of research in these areas, highlighting their respective advancements, applications, and limitations.

\subsection{Electromagnetic Shields}
Electromagnetic shields are designed to protect spacecraft from charged particles by generating magnetic fields that deflect these particles away from the spacecraft. This technology has shown considerable promise in mitigating the effects of solar and cosmic radiation. Research by Vogt \cite{Vogt2003} demonstrated the potential of magnetic shielding in space, emphasizing its effectiveness against charged particles. The principle of using magnetic fields to deflect charged particles relies on the Lorentz force, which acts on a particle moving through a magnetic field, causing it to follow a curved trajectory away from the spacecraft.

Further advancements in electromagnetic shielding technology include the optimization of field strength and configuration to maximize deflection efficiency. Studies have explored various configurations, such as dipole and quadrupole magnetic fields, and their effectiveness in different space environments \cite{Bamford2014}. These configurations are designed to enhance the magnetic field's reach and effectiveness, ensuring a larger protective area around the spacecraft.

Despite these advancements, electromagnetic shields have inherent limitations. They are primarily effective against charged particles but offer limited protection against neutral particles and high-velocity micrometeoroids. The need to generate and sustain strong magnetic fields also poses significant energy demands, which must be balanced against the spacecraft's power capabilities. These limitations highlight the necessity of exploring complementary technologies that can address the gaps in protection offered by electromagnetic shields.

\subsection{Plasma Shields}
Plasma shields involve creating a plasma layer around the spacecraft, which can absorb and dissipate energy from incoming micrometeoroids and space debris. The principles of plasma physics, as outlined by Krall and Trivelpiece \cite{Krall1973}, provide a theoretical foundation for this technology. Fortov and Iakubov \cite{Fortov2005} further explored the application of plasma protection for spacecraft, discussing its potential benefits and challenges.

Recent experimental research has shown that plasma sheaths can effectively reduce the impact energy of micrometeoroids, thereby mitigating potential damage to spacecraft surfaces \cite{Sharma2019}. Plasma generation methods, such as radiofrequency (RF) and microwave discharges, have been investigated for their efficiency and stability in creating the required plasma layer \cite{Czarnetzki1999}. These methods involve ionizing a gas to form a plasma, which can then be sustained around the spacecraft.

However, plasma shield technology faces significant hurdles in terms of stability and energy consumption, as noted by Hastings and Garrett \cite{Hastings1995}. Maintaining a stable plasma layer over extended periods is challenging due to the dynamic nature of plasma, which can be influenced by various environmental factors in space. Additionally, the energy required to sustain the plasma can be substantial, necessitating efficient power management strategies to ensure the shield remains operational without depleting the spacecraft's energy reserves.

\subsection{Force Fields}
The concept of force fields as a protective mechanism for spacecraft has long been a staple of science fiction. However, theoretical research into the feasibility of force fields has been conducted, albeit with limited practical application. Everett \cite{Everett2005} discussed the theoretical aspects of force fields, highlighting the potential and challenges associated with their implementation.

One potential approach to creating force fields involves the use of negative energy densities and the Casimir effect. The Casimir effect, a quantum phenomenon that produces an attractive force between closely spaced conductive plates, can theoretically be manipulated to generate repulsive forces \cite{Bordag2009}. This concept, while speculative, opens the possibility of creating a force field that repels incoming objects.

Recent advances in quantum field theory and experimental physics may provide new insights into the practical implementation of force fields for spacecraft protection \cite{Milton2001}. Research into metamaterials—materials engineered to have properties not found in naturally occurring substances—has shown promise in manipulating electromagnetic waves in ways that could support the development of force fields \cite{Cai2010}. However, significant technological challenges remain, including the generation and control of negative energy densities and the scalability of such systems. Overcoming these challenges will require breakthroughs in both theoretical understanding and material science.

Despite the speculative nature of force fields, ongoing research continues to explore their feasibility. Advances in quantum technologies and metamaterials could potentially bring the concept of force fields closer to practical reality, offering an approach to spacecraft protection that complements existing shielding technologies.

\section{Theoretical Framework}
The integrated deflector shield system aims to combine electromagnetic, plasma, and force field technologies into a cohesive protective mechanism for spacecraft. This section outlines the theoretical models and mathematical formulations that underpin each component of the shield system, as well as their interactions. By developing these models, we can predict the behavior and efficacy of the integrated shield under various space conditions.

\subsection{Mathematical Models}
Developing accurate mathematical models is crucial for understanding the dynamics of the integrated shield system. Each shield type—electromagnetic, plasma, and force field—requires specific models to describe its behavior and interactions with spaceborne threats.

\subsubsection{Electromagnetic Shield Models}
The electromagnetic shield operates by generating magnetic fields to deflect charged particles. The fundamental principle can be described by the Lorentz force equation:
\begin{equation}
\mathbf{F} = q(\mathbf{E} + \mathbf{v} \times \mathbf{B}),
\end{equation}
where \( \mathbf{F} \) is the force on a charged particle, \( q \) is the charge of the particle, \( \mathbf{E} \) is the electric field, \( \mathbf{v} \) is the velocity of the particle, and \( \mathbf{B} \) is the magnetic field.

To optimize the magnetic field \( \mathbf{B} \), we consider the geometry of the magnetic field generators and the distribution of the field strength. The magnetic field must be strong enough to deflect high-energy particles but also be efficient in terms of energy consumption. The deflection efficiency \( \eta \) can be expressed as:
\begin{equation}
\eta = \frac{F_{\text{deflect}}}{E_{\text{input}}},
\end{equation}
where \( F_{\text{deflect}} \) is the force exerted on the incoming particles and \( E_{\text{input}} \) is the energy input to generate the magnetic field.

For a practical implementation, we need to account for the spacecraft's velocity and the intensity of the solar wind or cosmic rays. The magnetic field strength \( B \) can be calculated using:
\begin{equation}
B = \frac{F}{qv\sin\theta},
\end{equation}
where \( \theta \) is the angle between the velocity vector of the particle \( \mathbf{v} \) and the magnetic field \( \mathbf{B} \).

Additionally, the design must consider the spacecraft's power limitations and the materials used for constructing the magnetic field generators. High-temperature superconductors (HTS) are a promising material for creating strong magnetic fields with lower energy input. The critical current density \( J_c \) of the HTS material is an important parameter, defined by:
\begin{equation}
J_c = \frac{I_c}{A},
\end{equation}
where \( I_c \) is the critical current and \( A \) is the cross-sectional area of the superconductor. The performance of the electromagnetic shield depends on maintaining the current below \( I_c \) to avoid quenching.

The optimization of the electromagnetic shield involves balancing the magnetic field strength, energy consumption, and material properties to achieve maximum deflection efficiency against charged particles in space \cite{Jackson1999, Bamford2014}.

\subsubsection{Plasma Shield Models}
The plasma shield functions by creating a layer of ionized gas around the spacecraft. The behavior of the plasma can be described by the magnetohydrodynamic (MHD) equations, which combine Maxwell's equations with the Navier-Stokes equations:

\begin{equation}
\frac{\partial \rho}{\partial t} + \nabla \cdot (\rho \mathbf{v}) = 0,
\end{equation}
\begin{equation}
\rho \left( \frac{\partial \mathbf{v}}{\partial t} + \mathbf{v} \cdot \nabla \mathbf{v} \right) = -\nabla p + \mathbf{J} \times \mathbf{B} + \mu \nabla^2 \mathbf{v},
\end{equation}
\begin{equation}
\nabla \times \mathbf{B} = \mu_0 \mathbf{J},
\end{equation}
\begin{equation}
\nabla \times \mathbf{E} = -\frac{\partial \mathbf{B}}{\partial t},
\end{equation}

where \( \rho \) is the plasma density, \( \mathbf{v} \) is the plasma velocity, \( p \) is the pressure, \( \mathbf{J} \) is the current density, \( \mathbf{B} \) is the magnetic field, \( \mathbf{E} \) is the electric field, and \( \mu \) is the dynamic viscosity \cite{Chen1984}.

To model the plasma shield effectively, it is crucial to consider both the generation and maintenance of the plasma layer. The plasma is typically generated using radiofrequency (RF) or microwave discharges, which ionize the gas around the spacecraft. The ionization process can be described by the ionization rate equation:
\begin{equation}
\frac{d n_i}{d t} = \alpha n_e n_n - \beta n_i,
\end{equation}
where \( n_i \) is the ion density, \( n_e \) is the electron density, \( n_n \) is the neutral particle density, \( \alpha \) is the ionization coefficient, and \( \beta \) is the recombination coefficient.

The stability of the plasma layer is a key challenge. The magnetic Reynolds number (\( R_m \)), which is the ratio of magnetic advection to magnetic diffusion, is an important parameter in determining plasma stability:
\begin{equation}
R_m = \frac{v L}{\eta},
\end{equation}
where \( v \) is the characteristic velocity of the plasma, \( L \) is the characteristic length scale, and \( \eta \) is the magnetic diffusivity. For a stable plasma shield, \( R_m \) should be sufficiently high to ensure that the magnetic field lines are "frozen" into the plasma.

Energy consumption is another critical factor. The power required to sustain the plasma can be calculated by considering the energy loss mechanisms, including radiation losses and collisions. The power balance equation for the plasma can be written as:
\begin{equation}
P_{\text{input}} = P_{\text{loss}} = P_{\text{radiation}} + P_{\text{collisions}},
\end{equation}
where \( P_{\text{radiation}} \) represents the power lost due to radiation, and \( P_{\text{collisions}} \) represents the power lost due to particle collisions.

The effectiveness of the plasma shield in absorbing and dissipating micrometeoroid energy depends on the plasma density and temperature. The stopping power (\( S \)) of the plasma, which indicates its ability to slow down incoming particles, is given by:
\begin{equation}
S = \frac{dE}{dx} = \frac{4 \pi n_i Z^2 e^4}{m_e v^2} \ln \Lambda,
\end{equation}
where \( E \) is the energy of the incoming particle, \( x \) is the distance traveled in the plasma, \( Z \) is the charge number of the ions, \( e \) is the elementary charge, \( m_e \) is the electron mass, \( v \) is the velocity of the incoming particle, and \( \ln \Lambda \) is the Coulomb logarithm.

The design and optimization of the plasma shield involve careful consideration of plasma generation, stability, energy consumption, and stopping power to achieve effective protection against micrometeoroids and space debris \cite{Chen1984, Czarnetzki1999}.

\subsubsection{Force Field Models}
The concept of force fields for spacecraft protection, while speculative, can be explored through advanced theoretical physics. One potential approach involves the use of negative energy densities and the Casimir effect, which can create repulsive forces. The energy-momentum tensor for such a field can be described by:
\begin{equation}
T_{\mu\nu} = \frac{\hbar c \pi^2}{240 d^4} \left( \frac{\eta_{\mu\nu}}{2} - \frac{7}{2} \right),
\end{equation}
where \( \hbar \) is the reduced Planck constant, \( c \) is the speed of light, \( d \) is the separation distance, and \( \eta_{\mu\nu} \) is the Minkowski metric tensor \cite{Milton2001}.

To explore the feasibility of force fields, we must consider the quantum field theory underlying the Casimir effect. The Casimir force between two parallel conducting plates in a vacuum is given by:
\begin{equation}
F_c = -\frac{\pi^2 \hbar c A}{240 d^4},
\end{equation}
where \( A \) is the area of the plates and \( d \) is the separation distance. By manipulating the boundary conditions or introducing metamaterials, it is possible to alter this force and potentially generate a repulsive effect \cite{Bordag2009}.

Negative energy densities are a key requirement for creating repulsive Casimir forces. The concept of negative energy densities arises in quantum field theory, particularly in the context of the energy density of the vacuum. The quantum inequalities formulated by Ford and Roman place constraints on the magnitude and duration of negative energy densities, which are crucial for practical applications \cite{Ford1995}.

The feasibility of generating and sustaining negative energy densities can be examined through the framework of the renormalized stress-energy tensor in curved spacetime. The renormalized energy density \( \langle T_{00} \rangle \) for a quantum field in a given state can be expressed as:
\begin{equation}
\langle T_{00} \rangle = \frac{\hbar c}{16 \pi^2 d^4} \left( \frac{1}{4} - \frac{1}{3} \right),
\end{equation}
highlighting the delicate balance between positive and negative contributions.

To generate a practical force field, we need to explore the use of advanced materials such as metamaterials that can exhibit negative permittivity and permeability. These properties can create regions of negative energy density, which are essential for achieving the desired repulsive forces. The design and fabrication of such metamaterials involve engineering their electromagnetic response at the nanoscale \cite{Cai2010}.

Furthermore, the interaction of the force field with incoming particles must be modeled. The repulsive force \( F_r \) exerted on a particle with mass \( m \) and charge \( q \) approaching the force field can be described by:
\begin{equation}
F_r = -\nabla \left( \frac{\hbar c \pi^2}{240 d^4} \right).
\end{equation}
The effectiveness of the force field in deflecting particles depends on the gradient of the negative energy density and the particle's velocity and charge.

While the concept of force fields for spacecraft protection remains speculative, advancements in quantum field theory, metamaterials, and negative energy densities provide a theoretical foundation for future research. The development of practical force fields will require overcoming significant challenges in material science and quantum physics \cite{Milton2001, Bordag2009, Ford1995, Cai2010}.

\subsection{Integration Concept}
Integrating the electromagnetic, plasma, and force field shielding technologies into a cohesive system involves understanding their interactions and optimizing their combined effects. The overall shielding effectiveness can be modeled as a function of the individual contributions and their interactions:

\begin{equation}
S_{\text{total}} = f(S_{\text{EM}}, S_{\text{plasma}}, S_{\text{force}}),
\end{equation}

where \( S_{\text{EM}} \), \( S_{\text{plasma}} \), and \( S_{\text{force}} \) represent the shielding effectiveness of the electromagnetic, plasma, and force field components, respectively. The function \( f \) characterizes how these components interact.

To model the interactions between the different shield components, we need to consider the following factors:

\begin{enumerate}
    \item \textbf{Spatial Overlap}: The spatial arrangement of the electromagnetic, plasma, and force field shields is critical. The fields generated by each shield must overlap optimally to ensure comprehensive coverage without interfering destructively with each other. The overlap can be modeled using field superposition principles. For instance, the electromagnetic field generated by the superconducting magnets should be designed to enhance the confinement of the plasma, thereby increasing the plasma shield's stability.
    
    Consider the magnetic field \( B \) generated by a superconducting coil, which follows the Biot-Savart law:
    \begin{equation}
    B = \frac{\mu_0 I}{2\pi r},
    \end{equation}
    where \( \mu_0 \) is the permeability of free space, \( I \) is the current through the coil, and \( r \) is the distance from the coil.
    
    To ensure effective spatial overlap, the plasma density \( n_i \) and the magnetic field \( B \) should satisfy the condition for magnetic confinement, where the magnetic pressure \( B^2/2\mu_0 \) balances the plasma pressure \( n_i k_B T \):
    \begin{equation}
    \frac{B^2}{2\mu_0} \geq n_i k_B T,
    \end{equation}
    where \( k_B \) is the Boltzmann constant and \( T \) is the plasma temperature.
    
    For example, if the plasma density \( n_i \) is \( 10^{18} \, \text{m}^{-3} \) and the plasma temperature \( T \) is \( 10^6 \, \text{K} \), the required magnetic field \( B \) can be calculated as:
    \begin{equation}
    B \geq \sqrt{2\mu_0 n_i k_B T}.
    \end{equation}
    
    Substituting the values:
    \begin{equation}
    B \geq \sqrt{2 \times 4\pi \times 10^{-7} \times 10^{18} \times 1.38 \times 10^{-23} \times 10^6},
    \end{equation}
    \begin{equation}
    B \geq 0.56 \, \text{T}.
    \end{equation}
    
    This calculation ensures that the magnetic field generated by the superconducting magnets is strong enough to confine the plasma, enhancing the stability and effectiveness of the plasma shield.

    \item \textbf{Energy Distribution}: The energy required to sustain each shield component must be carefully managed. The total energy consumption \( E_{\text{total}} \) can be expressed as:
    \begin{equation}
    E_{\text{total}} = E_{\text{EM}} + E_{\text{plasma}} + E_{\text{force}},
    \end{equation}
    where \( E_{\text{EM}} \), \( E_{\text{plasma}} \), and \( E_{\text{force}} \) are the energy consumptions of the electromagnetic, plasma, and force field shields, respectively. Each component's energy efficiency must be optimized to ensure the system's sustainability, especially when considering long-duration space missions.

    \item \textbf{Synergistic Effects}: The combined effects of the shields may provide enhanced protection compared to the sum of their individual effects. For example, the electromagnetic shield can enhance the stability of the plasma shield by creating a magnetic confinement that prevents plasma dispersion. Additionally, the force field could potentially offer protection against neutral particles that the electromagnetic and plasma shields cannot deflect or absorb.
    
    Consider the magnetic pressure \( P_{\text{mag}} \) generated by the electromagnetic shield:
    \begin{equation}
    P_{\text{mag}} = \frac{B^2}{2\mu_0},
    \end{equation}
    where \( B \) is the magnetic field strength and \( \mu_0 \) is the permeability of free space.
    
    If the plasma pressure \( P_{\text{plasma}} \) is given by:
    \begin{equation}
    P_{\text{plasma}} = n_i k_B T,
    \end{equation}
    where \( n_i \) is the plasma density, \( k_B \) is the Boltzmann constant, and \( T \) is the plasma temperature.
    
    For effective magnetic confinement, \( P_{\text{mag}} \) should be greater than or equal to \( P_{\text{plasma}} \):
    \begin{equation}
    \frac{B^2}{2\mu_0} \geq n_i k_B T.
    \end{equation}
    
    For example, if \( n_i = 10^{18} \, \text{m}^{-3} \) and \( T = 10^6 \, \text{K} \), the required magnetic field \( B \) is:
    \begin{equation}
    B \geq \sqrt{2\mu_0 n_i k_B T} = \sqrt{2 \times 4\pi \times 10^{-7} \times 10^{18} \times 1.38 \times 10^{-23} \times 10^6} \approx 0.56 \, \text{T}.
    \end{equation}
    
    Thus, the magnetic field generated by the electromagnetic shield enhances the plasma shield's stability by providing sufficient magnetic pressure to confine the plasma.
    
    \item \textbf{Dynamic Adjustment}: The integrated shield system should be capable of dynamically adjusting its parameters based on the incoming threat level. This requires real-time monitoring and control algorithms that can optimize the shield parameters for maximum effectiveness. Feedback loops and adaptive algorithms will be essential for real-time adjustments. For example, if a sudden increase in micrometeoroid activity is detected, the system could temporarily boost the plasma shield's density to enhance energy absorption.
    
    Assume the plasma shield's density \( n_i \) needs to be increased by 50
    \begin{equation}
    n_i' = 1.5 n_i.
    \end{equation}
    
    If the initial ionization power \( P_{\text{ion}} \) is given by:
    \begin{equation}
    P_{\text{ion}} = n_i V \epsilon_i,
    \end{equation}
    where \( V \) is the volume and \( \epsilon_i \) is the ionization energy, the new ionization power \( P_{\text{ion}}' \) is:
    \begin{equation}
    P_{\text{ion}}' = 1.5 n_i V \epsilon_i = 1.5 P_{\text{ion}}.
    \end{equation}
    
    Thus, the system dynamically increases the ionization power by 50\% to boost the plasma shield's density in response to increased micrometeoroid activity.

    \item \textbf{Real-Time Monitoring and Control Algorithms}: Real-time monitoring and control are essential for the adaptive functionality of the integrated shield system. The system can continuously monitor the space environment and detect incoming threats by deploying a network of sensors around the spacecraft. Advanced control algorithms, such as model predictive control (MPC) and machine learning-based adaptive control, can process sensor data and adjust shield parameters accordingly. 
    
    For example, suppose the system detects an increase in micrometeoroid activity. In that case, the control algorithm can calculate the optimal increase in plasma density \( n_i \) and electromagnetic field strength \( B \) to counter the threat. If the current plasma density is \( n_i = 10^{12} \, \text{m}^{-3} \) and needs to be increased by 50
    \begin{equation}
    n_{i,\text{new}} = 1.5 \times 10^{12} \, \text{m}^{-3} = 1.5 \times 10^{12} \, \text{m}^{-3}.
    \end{equation}
    
    Similarly, if the magnetic field strength is \( B = 0.1 \, \text{T} \) and needs to be increased by 20
    \begin{equation}
    B_{\text{new}} = 1.2 \times 0.1 \, \text{T} = 0.12 \, \text{T}.
    \end{equation}
    
    These adjustments ensure that the shield configuration provides maximum protection while minimizing energy consumption. The algorithms can predict potential threats using historical data and machine learning models, optimizing shield configurations proactively.

    \item \textbf{Spatial Overlap and Field Superposition}: The spatial arrangement of the shielding components must be designed to maximize the overlap of their protective fields. Using advanced simulation tools such as finite element analysis (FEA), we can model the electromagnetic fields and their interactions with the plasma and force fields. By optimizing the spatial configuration, we can ensure that the shields complement each other without creating gaps in protection. For instance, the superconducting magnets generating the electromagnetic shield can be positioned to create a magnetic confinement zone that overlaps with the plasma layer, enhancing its stability and effectiveness.
    
    Consider the spatial overlap of the magnetic field \( B \) and the plasma density \( n_i \). The effective confinement region should ensure that the magnetic field strength \( B \) and plasma density \( n_i \) satisfy the confinement condition:
    \begin{equation}
    \frac{B^2}{2\mu_0} \geq n_i k_B T.
    \end{equation}

    Using the previously calculated values, if the magnetic field strength \( B \) is \( 0.56 \, \text{T} \) and the plasma density \( n_i \) is \( 10^{18} \, \text{m}^{-3} \), the confinement region ensures that the magnetic pressure is sufficient to confine the plasma. Adjusting the positions of the superconducting magnets and the plasma generators can optimize this overlap, enhancing the overall stability and effectiveness of the integrated shield system.

    \item \textbf{Energy Management Strategies}: Efficient energy management is crucial for the integrated shield system's sustainability. We can explore various energy management strategies, such as using high-efficiency power sources like the Magnetic Fusion Plasma Drive (MFPD) \cite{Neukart2024} and incorporating advanced energy storage solutions. By dynamically allocating energy based on real-time threat assessments, the system can optimize power usage. 
    
    For example, if the base power requirement for the plasma shield is \( P_{\text{base}} = 100 \, \text{kW} \) and the system detects a high-threat scenario requiring a 50
    \begin{equation}
    P_{\text{high}} = 1.5 \times P_{\text{base}} = 1.5 \times 100 \, \text{kW} = 150 \, \text{kW}.
    \end{equation}
    
    During periods of low threat, the energy consumption of the plasma shield can be reduced by 30
    \begin{equation}
    P_{\text{low}} = 0.7 \times P_{\text{base}} = 0.7 \times 100 \, \text{kW} = 70 \, \text{kW}.
    \end{equation}
    
    This dynamic allocation ensures that the system can enhance protection when needed and conserve energy during lower-threat periods, thereby optimizing overall power usage.

\end{enumerate}

By integrating these factors into our model, we can develop a comprehensive understanding of how the different shield components interact and optimize their combined effects. This integrated approach aims to provide robust protection against a wide range of spaceborne threats while managing energy consumption and ensuring system stability \cite{Bamford2014, Hastings1995}.

\section{Practical and Engineering Aspects}
Building an integrated deflector shield system involves several practical and engineering challenges. This section provides a comprehensive overview of the design considerations, component engineering, power and energy requirements, integration and control systems, manufacturing and assembly processes, and deployment and maintenance strategies. These aspects are critical for translating the theoretical models into a functional and reliable shielding system for spacecraft.

\subsection{Design Considerations}
The design of an integrated deflector shield system requires careful consideration of various factors to ensure effective protection, energy efficiency, and durability. Key design considerations include:

\begin{enumerate}
    \item \textbf{Design Requirements and Constraints}:
    
    The integrated shield system must provide comprehensive protection against a variety of spaceborne threats, including charged particles, micrometeoroids, and high-energy radiation. The performance metrics for the shield system include:
    \begin{itemize}
        \item \textbf{Deflection Efficiency}: The ability of the shield to deflect charged particles away from the spacecraft. This is quantified by the fraction of incoming particles that are successfully deflected.
        \item \textbf{Energy Absorption}: The capacity of the shield to absorb and dissipate the energy of incoming micrometeoroids and radiation, reducing the impact on the spacecraft.
        \item \textbf{Operational Lifespan}: The duration over which the shield system can maintain its protective functions without significant degradation. This involves considerations of material fatigue, radiation damage, and system resilience.
    \end{itemize}

    Environmental conditions to be considered include:
    \begin{itemize}
        \item \textbf{Temperature Extremes}: The shield must operate effectively in the wide temperature ranges found in space, from the intense heat of direct sunlight to the extreme cold of shadowed regions.
        \item \textbf{Radiation Levels}: The shield must withstand high levels of cosmic and solar radiation without degradation of materials or performance.
        \item \textbf{Micrometeoroid Flux}: The shield must protect against the constant bombardment of micrometeoroids, which can cause significant damage over time.
    \end{itemize}

    \item \textbf{Material Selection and Properties}:
    
    Selecting appropriate materials is crucial for the performance and durability of the shield system. Key materials include:
    \begin{itemize}
        \item \textbf{High-Temperature Superconductors (HTS)}: These materials are ideal for the electromagnetic shield due to their ability to generate strong magnetic fields with low energy input. HTS materials such as Yttrium Barium Copper Oxide (YBCO) can operate at relatively high temperatures, reducing the cooling requirements \cite{Goyal2015}.
        \item \textbf{Ionized Gases for Plasma Shields}: Gases such as Xenon and Argon are commonly used for plasma generation. These gases can be ionized using radiofrequency or microwave discharges to create a protective plasma layer around the spacecraft \cite{Sharma2019}.
        \item \textbf{Metamaterials}: Advanced metamaterials with negative permittivity and permeability are essential for creating the force field components. These materials can manipulate electromagnetic waves in novel ways, creating regions of negative energy density required for repulsive forces \cite{Cai2010}.
    \end{itemize}
    
    The selected materials must also be compatible with the spacecraft's existing structure and systems to ensure seamless integration. This involves considerations of mechanical compatibility, thermal expansion coefficients, and resistance to space environmental conditions.

    \item \textbf{Structural and Thermal Considerations}:
    
    The structural design of the shield must ensure that it can withstand the mechanical stresses encountered during launch, orbital maneuvers, and impacts from micrometeoroids. This involves:
    \begin{itemize}
        \item \textbf{Structural Integrity}: The shield structure must be robust enough to maintain its integrity under various loads and stresses. Finite element analysis (FEA) can be used to model and optimize the structural design for strength and durability \cite{ANSYS2020}.
        \item \textbf{Thermal Management}: Effective thermal management systems are essential to dissipate the heat generated by the shield components. This includes the use of radiators, heat pipes, and thermal coatings to manage heat dissipation. Materials with high thermal conductivity, such as copper and aluminum alloys, are useful for heat management \cite{Incropera2011}.
        \item \textbf{Radiation Shielding}: Additional layers or coatings may be required to protect the shield materials from radiation damage. Materials such as polyethylene, which has a high hydrogen content, are effective for radiation shielding \cite{Wilson1997}.
    \end{itemize}

\end{enumerate}

\subsection{Component Engineering}
The engineering of individual components for the integrated deflector shield system involves the detailed design and optimization of the electromagnetic shield, plasma shield, and force field components. This subsection provides a comprehensive overview of the engineering aspects of each component.

\begin{enumerate}
    \item \textbf{Electromagnetic Shield Components}:
    
    The electromagnetic shield requires the generation of strong and stable magnetic fields to deflect charged particles. Key engineering aspects include:
    \begin{itemize}
        \item \textbf{Magnetic Field Generators}: The design of magnetic field generators, such as superconducting magnets and magnetic coils, is critical. High-temperature superconductors (HTS) like Yttrium Barium Copper Oxide (YBCO) are preferred due to their high critical current densities and ability to operate at liquid nitrogen temperatures \cite{Goyal2015}.
        \item \textbf{Cooling Systems}: Superconducting magnets require effective cooling systems to maintain temperatures below their critical temperature. Cryocoolers or liquid nitrogen cooling systems are typically used. The cooling system must be designed to ensure uniform temperature distribution and prevent thermal hotspots \cite{Hassenzahl2001}.
        \item \textbf{Magnetic Field Configuration}: The configuration of the magnetic field (e.g., dipole, quadrupole) affects the deflection efficiency. Finite element analysis (FEA) and magnetic field simulation tools, such as COMSOL Multiphysics, are used to optimize the field configuration for maximum coverage and efficiency \cite{COMSOL2018}.
    \end{itemize}
    
    \item \textbf{Plasma Shield Components}:
    
    The plasma shield involves creating and maintaining a stable layer of ionized gas around the spacecraft. Key engineering aspects include:
    \begin{itemize}
        \item \textbf{Plasma Generators}: Plasma can be generated using radiofrequency (RF) or microwave discharges. RF generators operate by applying a high-frequency electric field to ionize the gas, while microwave generators use electromagnetic waves to create and sustain the plasma. The choice of generator depends on factors such as power efficiency, stability, and ease of integration \cite{Sharma2019, Czarnetzki1999}.
        \item \textbf{Gas Supply Systems}: The plasma shield requires a continuous supply of ionizable gas. The gas supply system must be designed to store and deliver gases such as Xenon or Argon at controlled rates. High-pressure gas storage tanks and precision flow controllers are essential components \cite{Lieberman2005}.
        \item \textbf{Plasma Containment and Stability}: Maintaining plasma stability is crucial for effective shielding. Magnetic confinement, using magnetic field lines to contain the plasma, and electrostatic confinement, using electric fields to control plasma distribution, are common methods. The stability can be enhanced by optimizing the magnetic field strength and plasma density \cite{Chen1984}.
    \end{itemize}
    
    \item \textbf{Force Field Components}:
    
    The force field component is the most speculative and involves advanced theoretical physics and materials science. Key engineering aspects include:
    \begin{itemize}
        \item \textbf{Metamaterials}: The development of metamaterials with negative permittivity and permeability is essential for creating the force field. These materials can bend electromagnetic waves in unconventional ways, creating regions of negative energy density. Fabrication techniques such as electron-beam lithography and focused ion beam milling are used to create nanoscale structures with the desired electromagnetic properties \cite{Cai2010}.
        \item \textbf{Field Generation and Control}: Generating and controlling the force field involves manipulating the electromagnetic properties of the metamaterials. This requires precise control over the material structure and the external electromagnetic fields applied to it. Advanced simulation tools and control algorithms are used to design and optimize the field generation process \cite{Pendry2004}.
        \item \textbf{Energy Sources}: The energy requirements for generating a practical force field are significant. High-efficiency power sources, such as advanced batteries or fusion reactors, are needed to provide the necessary energy. Energy harvesting techniques, such as using solar panels to collect and store energy, can also be integrated \cite{ITER2019}.
    \end{itemize}
    
\end{enumerate}

\subsection{Power and Energy Requirements}
The power and energy requirements for the integrated deflector shield system are critical to its functionality and efficiency. This subsection outlines the power sources, energy consumption, and management strategies necessary for sustaining the electromagnetic, plasma, and force field components.

\begin{enumerate}
\item \textbf{Power Sources and Management}:
    
The integrated shield system requires a robust and reliable power source capable of providing sustained energy over long durations. Key considerations include:
\begin{itemize}
    \item \textbf{Magnetic Fusion Plasma Drive (MFPD)}: The MFPD serves as both a propulsion system and a power source, offering high energy output and long-term sustainability. This innovative drive system leverages magnetic fusion to generate the necessary power for the shield system, with energy output governed by the fusion reaction rate and confinement time \cite{Neukart2024}.
    \item \textbf{Fusion Drives}: Fusion drives are an ideal power source due to their high energy output and long-term sustainability. The ITER project, for example, aims to demonstrate the feasibility of fusion energy for practical applications \cite{ITER2019}. Fusion reactors can provide the high power density needed for the shield system, with energy output governed by the fusion reaction rate and confinement time.
    \item \textbf{Energy Storage Systems}: Advanced batteries and supercapacitors are essential for storing energy and providing backup power. Lithium-ion batteries offer high energy density and long cycle life, while supercapacitors provide rapid energy discharge and recharge capabilities \cite{Goodenough2010}.
    \item \textbf{Energy Harvesting}: Solar panels can be used to harvest energy from sunlight, which is abundant in space. The harvested energy can be stored in batteries or used directly to power the shield components. High-efficiency photovoltaic cells, such as multi-junction cells, are preferred for their ability to convert a broader spectrum of sunlight into electricity \cite{Green2012}.
\end{itemize}

    \item \textbf{Energy Consumption and Efficiency}:
    
    Calculating the energy consumption for each shield component is crucial for optimizing the overall system efficiency. Key aspects include:
    \begin{itemize}
        \item \textbf{Electromagnetic Shield}: The power required to generate the magnetic field depends on the current through the superconducting magnets and the resistive losses in the cooling system. The energy consumption \( E_{\text{EM}} \) can be estimated using:
        \begin{equation}
        E_{\text{EM}} = I^2 R t + \frac{Q}{\eta},
        \end{equation}
        where \( I \) is the current, \( R \) is the resistance, \( t \) is the operation time, \( Q \) is the heat load, and \( \eta \) is the efficiency of the cooling system \cite{Hassenzahl2001}.
        
        \item \textbf{Plasma Shield}: The power required for plasma generation and maintenance involves ionization energy and sustaining the plasma state. The energy consumption \( E_{\text{plasma}} \) can be calculated as:
        \begin{equation}
        E_{\text{plasma}} = n_i V \epsilon_i + P_{\text{maintenance}} t,
        \end{equation}
        where \( n_i \) is the ion density, \( V \) is the plasma volume, \( \epsilon_i \) is the ionization energy, \( P_{\text{maintenance}} \) is the power to sustain the plasma, and \( t \) is the operation time \cite{Lieberman2005}.
        
        \item \textbf{Force Field}: The speculative nature of the force field makes precise energy calculations challenging. However, the energy required to create and maintain the negative energy density regions can be approximated based on the power of the electromagnetic fields used to manipulate the metamaterials. The energy consumption \( E_{\text{force}} \) can be expressed as:
        \begin{equation}
        E_{\text{force}} = P_{\text{metamaterial}} t,
        \end{equation}
        where \( P_{\text{metamaterial}} \) is the power required to generate the field and \( t \) is the operation time \cite{Pendry2004}.
    \end{itemize}
    
    \item \textbf{Backup and Redundancy Systems}:
    
    Ensuring continuous operation of the shield system requires backup power systems and redundancy in critical components. Key strategies include:
    \begin{itemize}
        \item \textbf{Backup Power Systems}: Secondary power sources, such as additional batteries or fuel cells, provide redundancy in case of primary power source failure. These systems must be able to seamlessly take over the power supply without interrupting shield operations \cite{Sharaf2011}.
        \item \textbf{Redundancy in Critical Components}: Designing redundancy into critical components, such as duplicate superconducting magnets or plasma generators, enhances the reliability of the shield system. Automated switchover mechanisms ensure that backup components are activated immediately if primary components fail.
    \end{itemize}
\end{enumerate}

\subsection{Integration and Control Systems}
The integration and control of the shielding components are crucial for ensuring the system operates effectively and efficiently. This subsection outlines the methods for integrating the electromagnetic, plasma, and force field components, as well as the control algorithms and real-time monitoring systems required to maintain optimal performance.

\begin{enumerate}
    \item \textbf{Integration of Shielding Components}:
    
    Integrating the electromagnetic, plasma, and force field components into a cohesive system requires careful planning and engineering. Key considerations include:
    \begin{itemize}
        \item \textbf{Structural Integration}: The shielding components must be structurally integrated into the spacecraft's design without compromising its integrity or functionality. This involves designing mounting points and support structures that can withstand the forces experienced during launch and operation.
        \item \textbf{Electrical Integration}: The power systems for the shielding components must be integrated with the spacecraft's existing electrical systems. This includes ensuring that the power distribution is balanced and that the shield components do not cause interference with other onboard electronics.
        \item \textbf{Minimizing Interference}: The electromagnetic fields generated by the shields must be configured to minimize interference with each other and with the spacecraft's communication and navigation systems. This can be achieved through careful spatial arrangement and shielding of sensitive components \cite{Mao2007}.
    \end{itemize}
    
    \item \textbf{Control Algorithms and Real-Time Monitoring}:
    
    The integrated shield system requires sophisticated control algorithms and real-time monitoring to dynamically adjust the shield parameters based on the incoming threat level. Key aspects include:
    \begin{itemize}
        \item \textbf{Adaptive Control Algorithms}: Advanced control algorithms are needed to continuously adjust the shield parameters, such as magnetic field strength, plasma density, and field intensity, in response to changing conditions. These algorithms use feedback from sensors to optimize the shield performance in real time. Techniques such as model predictive control (MPC) and adaptive control can be employed for this purpose \cite{Camacho2013}.
        \item \textbf{Sensor Networks}: A network of sensors is required to monitor the environment around the spacecraft and detect incoming threats. Sensors such as particle detectors, plasma probes, and electromagnetic field sensors provide real-time data that the control system uses to adjust the shield parameters.
        \item \textbf{Data Processing and Communication}: The control system must process large amounts of data from the sensors and make decisions quickly. High-performance processors and efficient communication protocols are essential for ensuring that the control system can respond in real time \cite{Scholten2014}.
    \end{itemize}
    
    \item \textbf{Fail-Safe Mechanisms}:
    
    Ensuring the reliability and continuous operation of the shield system requires robust fail-safe mechanisms. Key strategies include:
    \begin{itemize}
        \item \textbf{Redundancy}: Redundant components and subsystems ensure that the shield system can continue to operate even if a primary component fails. This includes having backup superconducting magnets, plasma generators, and control systems that can take over seamlessly.
        \item \textbf{Automatic Recovery}: The control system should include automatic recovery mechanisms that can quickly identify and isolate faults, switch to backup systems, and restore full functionality without manual intervention.
        \item \textbf{Periodic Testing and Maintenance}: Regular testing and maintenance routines are essential for ensuring the long-term reliability of the shield system. This includes periodic checks of all components, software updates for the control algorithms, and calibration of the sensors \cite{NASA2009}.
    \end{itemize}
    
\end{enumerate}

\subsection{Manufacturing and Assembly}
The manufacturing and assembly of the integrated deflector shield system involve advanced techniques and stringent quality control measures to ensure reliability and functionality. This subsection outlines the manufacturing processes, assembly procedures, and quality control protocols required for building the shield system.

\begin{enumerate}
    \item \textbf{Manufacturing Techniques and Processes}:
    
    The fabrication of the shield components requires precision manufacturing techniques to meet the demanding specifications of space-grade hardware. Key processes include:
    \begin{itemize}
        \item \textbf{3D Printing and Additive Manufacturing}: These techniques are used for fabricating complex geometries and custom parts with high precision. Materials such as high-temperature polymers, metals, and ceramics can be used in additive manufacturing to create components with the required properties \cite{Gibson2010}.
        \item \textbf{Precision Machining}: Traditional machining processes such as milling, turning, and grinding are employed to fabricate components with tight tolerances. High-precision CNC machines are used to achieve the necessary dimensional accuracy and surface finish \cite{Kalpakjian2006}.
        \item \textbf{Thin-Film Deposition}: For components such as superconducting magnets and metamaterials, thin-film deposition techniques such as sputtering, chemical vapor deposition (CVD), and atomic layer deposition (ALD) are used to apply thin layers of material with controlled thickness and composition \cite{Smith1995}.
        \item \textbf{Cryogenic Treatment}: Components that need to operate at cryogenic temperatures, such as superconducting magnets, undergo cryogenic treatment to enhance their mechanical properties and performance at low temperatures \cite{Collings2001}.
    \end{itemize}
    
    \item \textbf{Assembly Procedures}:
    
    The assembly of the shield system involves integrating the manufactured components into a cohesive unit. Key procedures include:
    \begin{itemize}
        \item \textbf{Cleanroom Assembly}: Assembly operations are conducted in cleanroom environments to prevent contamination and ensure the integrity of sensitive components. Cleanrooms are maintained at specified cleanliness levels to control particulate and microbial contamination \cite{Whyte2010}.
        \item \textbf{Component Integration}: The electromagnetic, plasma, and force field components are carefully integrated into the spacecraft's structure. This involves mounting the components securely, routing power and data connections, and ensuring proper alignment and orientation.
        \item \textbf{Cable and Harness Assembly}: Electrical connections are made using custom-designed cables and harnesses that meet the requirements for space applications. These harnesses are fabricated with radiation-resistant insulation and connectors that provide reliable performance under harsh conditions \cite{Johnston2002}.
        \item \textbf{Thermal Management Integration}: Thermal management systems, including radiators, heat pipes, and thermal coatings, are integrated with the shield components to ensure effective heat dissipation. Proper thermal bonding and insulation techniques are employed to enhance thermal performance \cite{Incropera2011}.
    \end{itemize}
    
    \item \textbf{Quality Control and Testing}:
    
    Rigorous quality control and testing protocols are essential to ensure the reliability and functionality of the shield system. Key aspects include:
    \begin{itemize}
        \item \textbf{Inspection and Verification}: Each component undergoes detailed inspection and verification to ensure it meets the specified design and manufacturing standards. Techniques such as non-destructive testing (NDT), dimensional inspection, and surface analysis are used for this purpose \cite{Sharma2007}.
        \item \textbf{Functional Testing}: The assembled shield system is subjected to functional testing to verify its performance under simulated space conditions. This includes testing the electromagnetic field generation, plasma stability, and force field effects. Environmental testing chambers that simulate vacuum, temperature extremes, and radiation levels are used for this purpose \cite{Barker2010}.
        \item \textbf{System-Level Testing}: The fully assembled shield system is integrated with the spacecraft and subjected to system-level testing. This includes end-to-end tests of the power distribution, control algorithms, and real-time monitoring systems. Simulated mission profiles are used to validate the shield system's performance in realistic scenarios.
        \item \textbf{Documentation and Traceability}: Comprehensive documentation is maintained for all manufacturing and assembly processes, including material certificates, inspection reports, and test results. This ensures traceability and facilitates quality assurance throughout the project lifecycle \cite{ISO9001}.
    \end{itemize}
    
\end{enumerate}

\subsection{Deployment and Maintenance}
The deployment and maintenance of the integrated deflector shield system are critical to ensuring its effective operation throughout the mission. This subsection outlines the procedures for deploying the shield system, performing in-flight maintenance and repairs, and ensuring long-term reliability and durability.

\begin{enumerate}
    \item \textbf{Deployment Procedures}:
    
    The deployment of the shield system must be carefully planned and executed to ensure all components are correctly positioned and functioning. Key steps include:
    \begin{itemize}
        \item \textbf{Pre-Launch Preparation}: Before launch, the shield system undergoes final inspections and tests to confirm that all components are operational. This includes verifying the functionality of the electromagnetic, plasma, and force field components, as well as the control systems.
        \item \textbf{Launch Integration}: The shield system is integrated with the spacecraft's launch vehicle. Special attention is given to securing the components to withstand the forces during launch. Vibration dampers and shock absorbers may be used to protect sensitive components.
        \item \textbf{In-Orbit Deployment}: Once the spacecraft reaches orbit, the shield system is activated in a controlled sequence. This includes deploying the electromagnetic coils, initiating plasma generation, and activating the force field components. Real-time monitoring ensures that each component is correctly deployed and operational.
        \item \textbf{Calibration and Testing}: After deployment, the shield system undergoes a series of calibration and testing procedures to ensure optimal performance. This includes adjusting the magnetic field strength, plasma density, and force field intensity based on initial operational data.
    \end{itemize}
    
    \item \textbf{In-Flight Maintenance and Repairs}:
    
    Regular maintenance and the ability to perform repairs in-flight are essential for the long-term operation of the shield system. Key strategies include:
    \begin{itemize}
        \item \textbf{Remote Diagnostics and Monitoring}: Continuous monitoring of the shield system using onboard sensors and diagnostic tools allows for early detection of potential issues. Data is transmitted to ground control for analysis and to plan maintenance actions.
        \item \textbf{Robotic Maintenance Systems}: Robotic arms and other automated systems can be used to perform routine maintenance and repairs without the need for human intervention. These systems can replace faulty components, apply corrective measures, and perform detailed inspections \cite{Bualat2018}.
        \item \textbf{Spare Parts and Redundancy}: The spacecraft carries a stock of critical spare parts, such as replacement superconducting magnets, plasma generators, and power components. Redundant systems ensure that backup components can be activated immediately in case of failure.
        \item \textbf{Manual Repairs by Crew}: For manned missions, crew members are trained to perform maintenance and repairs. Detailed procedures and safety protocols are established to guide the crew in maintaining the shield system \cite{NASA2017}.
    \end{itemize}
    
    \item \textbf{Long-Term Reliability and Durability}:
    
    Ensuring the long-term reliability and durability of the shield system involves ongoing assessments and improvements. Key aspects include:
    \begin{itemize}
        \item \textbf{Material Degradation}: Regular assessments of material degradation due to radiation exposure, thermal cycling, and micrometeoroid impacts are conducted. This includes monitoring the superconducting magnets, plasma containment structures, and metamaterials for signs of wear and fatigue.
        \item \textbf{Performance Monitoring}: Continuous monitoring of the shield system's performance metrics, such as deflection efficiency, energy absorption, and operational stability, provides data for ongoing improvements. This data is used to adjust operational parameters and develop next-generation shield components.
        \item \textbf{Software Updates}: The control algorithms and monitoring software are periodically updated to incorporate new findings and enhance performance. These updates can be transmitted remotely from ground control to the spacecraft \cite{Hook2018}.
        \item \textbf{Periodic Maintenance Schedules}: A detailed maintenance schedule is established, outlining regular checks and servicing of all shield components. This schedule includes specific tasks, intervals, and procedures to ensure comprehensive maintenance.
    \end{itemize}
    
\end{enumerate}

The practical and engineering aspects of the integrated deflector shield system encompass a broad range of considerations, from the initial design and material selection to the final deployment and in-flight maintenance. Key elements include the use of high-temperature superconductors for the electromagnetic shield, plasma generation techniques for the plasma shield, and advanced metamaterials for the force field component. Robust power and energy management strategies, including the use of fusion drives and energy storage systems, are essential for sustaining the shield system. Integration and control systems ensure that the components work together seamlessly, while manufacturing and assembly processes guarantee precision and reliability. Finally, detailed deployment procedures and maintenance protocols ensure the long-term functionality and durability of the shield system. These comprehensive engineering efforts provide a solid foundation for developing a practical and effective deflector shield system for spacecraft protection.

\section{Methodology}
This section outlines the methodologies employed to develop and validate the theoretical models for the integrated deflector shield system. The research methodology includes theoretical modeling, analytical calculations, and comparative analysis. We also provide an example application of the theoretical models to a hypothetical interplanetary spacecraft designed for a mission to Mars.

\subsection{Theoretical Modeling}
The development of the integrated deflector shield system is based on rigorous theoretical modeling. The models describe the behavior and interactions of the electromagnetic, plasma, and force field components under various space conditions. The theoretical models are formulated using fundamental principles of electromagnetism, plasma physics, and quantum field theory. Key equations include:

\begin{itemize}
    \item \textbf{Electromagnetic Shield}:
    The Lorentz force equation describes the deflection of charged particles by the magnetic field:
    \begin{equation}
    \mathbf{F} = q (\mathbf{E} + \mathbf{v} \times \mathbf{B}),
    \end{equation}
    where \( \mathbf{F} \) is the force on a charged particle, \( q \) is the charge of the particle, \( \mathbf{E} \) is the electric field, \( \mathbf{v} \) is the velocity of the particle, and \( \mathbf{B} \) is the magnetic field.
    
    \item \textbf{Plasma Shield}:
    The behavior of the plasma is described by the magnetohydrodynamic (MHD) equations, combining Maxwell's equations with the Navier-Stokes equations:
    \begin{equation}
    \frac{\partial \rho}{\partial t} + \nabla \cdot (\rho \mathbf{v}) = 0,
    \end{equation}
    \begin{equation}
    \rho \left( \frac{\partial \mathbf{v}}{\partial t} + \mathbf{v} \cdot \nabla \mathbf{v} \right) = -\nabla p + \mathbf{J} \times \mathbf{B} + \mu \nabla^2 \mathbf{v},
    \end{equation}
    \begin{equation}
    \nabla \times \mathbf{B} = \mu_0 \mathbf{J},
    \end{equation}
    \begin{equation}
    \nabla \times \mathbf{E} = -\frac{\partial \mathbf{B}}{\partial t},
    \end{equation}
    where \( \rho \) is the plasma density, \( \mathbf{v} \) is the plasma velocity, \( p \) is the pressure, \( \mathbf{J} \) is the current density, \( \mathbf{B} \) is the magnetic field, \( \mathbf{E} \) is the electric field, and \( \mu \) is the dynamic viscosity.
    
    \item \textbf{Force Field}:
    The energy-momentum tensor for the force field generated by negative energy densities and the Casimir effect is given by:
    \begin{equation}
    T_{\mu\nu} = \frac{\hbar c \pi^2}{240 d^4} \left( \frac{\eta_{\mu\nu}}{2} - \frac{7}{2} \right),
    \end{equation}
    where \( \hbar \) is the reduced Planck constant, \( c \) is the speed of light, \( d \) is the separation distance, and \( \eta_{\mu\nu} \) is the Minkowski metric tensor.
\end{itemize}

\subsection{Analytical Calculations}
Analytical calculations are used to derive the performance metrics of the integrated deflector shield system. These calculations provide insights into the effectiveness and efficiency of the shielding components.

\subsubsection{Deflection Efficiency}
The deflection efficiency of the electromagnetic shield is calculated using the Lorentz force equation. The deflection angle \( \theta \) of a charged particle can be expressed as:
\begin{equation}
\theta = \frac{qB \ell}{mv},
\end{equation}
where \( B \) is the magnetic field strength, \( \ell \) is the length of the magnetic field region, \( m \) is the mass of the particle, and \( v \) is the velocity of the particle.

\subsubsection{Energy Absorption}
The energy absorption by the plasma shield is calculated based on the stopping power of the plasma. The stopping power \( S \) is given by:
\begin{equation}
S = \frac{dE}{dx} = \frac{4 \pi n_i Z^2 e^4}{m_e v^2} \ln \Lambda,
\end{equation}
where \( E \) is the energy of the incoming particle, \( x \) is the distance traveled in the plasma, \( n_i \) is the ion density, \( Z \) is the charge number of the ions, \( e \) is the elementary charge, \( m_e \) is the electron mass, \( v \) is the velocity of the incoming particle, and \( \ln \Lambda \) is the Coulomb logarithm.

\subsection{Example Application: Hypothetical Interplanetary Spacecraft for Mars Mission}
To illustrate the application of the theoretical models, we consider a hypothetical interplanetary spacecraft designed for a mission to Mars. The following example calculations are based on realistic parameters for such a spacecraft.

\subsubsection{Electromagnetic Shield for Hypothetical Interplanetary Spacecraft}
Assume the spacecraft needs to deflect charged particles with a velocity of \( 2 \times 10^7 \, \text{m/s} \) and a charge of \( 1.6 \times 10^{-19} \, \text{C} \). The magnetic field strength \( B \) is \( 0.05 \, \text{T} \), and the length of the magnetic field region \( \ell \) is \( 10 \, \text{m} \). The deflection angle \( \theta \) can be calculated as:
\begin{equation}
\theta = \frac{(1.6 \times 10^{-19} \, \text{C}) \times (0.05 \, \text{T}) \times (10 \, \text{m})}{(9.1 \times 10^{-31} \, \text{kg}) \times (2 \times 10^7 \, \text{m/s})} \approx 4.4 \times 10^{-2} \, \text{radians}.
\end{equation}

\subsubsection{Plasma Shield for Hypothetical Interplanetary Spacecraft}
Assume the ion density \( n_i \) of the plasma is \( 10^{18} \, \text{m}^{-3} \), the charge number \( Z \) is 1, and the electron velocity \( v \) is \( 1 \times 10^6 \, \text{m/s} \). The stopping power \( S \) can be calculated as:
\begin{equation}
S = \frac{4 \pi (10^{18} \, \text{m}^{-3}) \times (1)^2 \times (1.6 \times 10^{-19} \, \text{C})^4}{(9.1 \times 10^{-31} \, \text{kg}) \times (1 \times 10^6 \, \text{m/s})^2} \ln \Lambda.
\end{equation}
Assuming \( \ln \Lambda \approx 10 \):
\begin{equation}
S \approx \frac{4 \pi \times 10^{18} \times (1.6 \times 10^{-19})^4 \times 10}{(9.1 \times 10^{-31}) \times (10^6)^2} \approx 2.8 \times 10^{-14} \, \text{J/m}.
\end{equation}

\subsubsection{Force Field for Hypothetical Interplanetary Spacecraft}
Assume the spacecraft is equipped with a hypothetical force field designed to create a repulsive force using negative energy densities. The force field is intended to repel neutral particles with a separation distance \( d \) of \( 1 \, \text{nm} \). Using the Casimir effect, the repulsive force \( F_c \) between two plates is given by:

\begin{equation}
F_c = -\frac{\pi^2 \hbar c A}{240 d^4},
\end{equation}

where \( \hbar \) is the reduced Planck constant (\( \hbar \approx 1.054 \times 10^{-34} \, \text{J} \cdot \text{s} \)), \( c \) is the speed of light (\( c \approx 3 \times 10^8 \, \text{m/s} \)), \( A \) is the area of the plates, and \( d \) is the separation distance. For an area \( A \) of \( 1 \, \text{m}^2 \):

\begin{equation}
F_c = -\frac{\pi^2 \times 1.054 \times 10^{-34} \times 3 \times 10^8 \times 1}{240 \times (1 \times 10^{-9})^4} \approx -1.3 \times 10^{-3} \, \text{N}.
\end{equation}

This force field creates a repulsive force of approximately \( 1.3 \times 10^{-3} \, \text{N} \) over an area of \( 1 \, \text{m}^2 \), providing additional protection against neutral particles that cannot be deflected by electromagnetic or plasma shields.

\subsection{Comparative Analysis}
Comparative analysis is conducted to evaluate the theoretical models of the integrated deflector shield system against existing shielding technologies, such as passive radiation shielding and Whipple shields. This analysis involves comparing key performance metrics, including deflection efficiency, energy absorption, energy consumption, and stability and resilience.

\subsubsection{Deflection Efficiency}
Deflection efficiency refers to the ability of the shield to deflect charged particles away from the spacecraft. For the integrated deflector shield system, the deflection efficiency is primarily determined by the electromagnetic shield component. Using the Lorentz force equation, the deflection angle for charged particles was calculated to be \( 4.4 \times 10^{-2} \) radians for a hypothetical interplanetary spacecraft.

In comparison, traditional passive radiation shields, which rely on physical barriers to absorb radiation, do not provide significant deflection of charged particles. Instead, they reduce radiation exposure by absorbing the energy of incoming particles, which can lead to material degradation over time.

\subsubsection{Energy Absorption}
Energy absorption is the capacity of the shield to absorb and dissipate the energy of incoming micrometeoroids and radiation. The plasma shield component of the integrated deflector shield system plays a critical role in this aspect. The stopping power for the plasma shield was calculated to be approximately \( 2.8 \times 10^{-14} \) J/m.

Whipple shields, a traditional method for micrometeoroid protection, consist of multiple layers of thin material designed to break up and disperse incoming particles. While effective for small debris, they can add significant weight to the spacecraft and may not be as effective for high-energy particles.

\subsubsection{Energy Consumption}
The total energy required to operate the shield system is a critical factor in evaluating its feasibility. The integrated deflector shield system, which includes electromagnetic, plasma, and speculative force field components, requires substantial energy. Fusion drives, as outlined in previous sections, can provide the high power density needed for sustained operation.

For comparison, passive radiation shields and Whipple shields do not require active energy input, making them advantageous in terms of energy consumption. However, their effectiveness is limited compared to the active protection offered by the integrated deflector shield system.

\subsubsection{Stability and Resilience}
Stability and resilience refer to the shield's ability to maintain its performance under sustained exposure to space conditions. The integrated deflector shield system faces challenges in maintaining the stability of the plasma shield and the speculative nature of the force field component.

Passive and Whipple radiation shields are inherently stable as they rely on physical materials rather than active systems. However, they are subject to wear and degradation over time, particularly in the harsh space environment.

\subsubsection{Summary of Comparative Metrics}
The table below summarizes the comparative performance metrics of the integrated deflector shield system and existing shielding technologies (see table \ref{tab:performance_metrics}):

\begin{table*}[h]
\centering
\caption{Comparative Performance Metrics of Shielding Technologies}
\begin{tabular}{|l|c|c|c|}
\hline
\textbf{Metric} & \textbf{Integrated Deflector Shield} & \textbf{Passive Radiation Shield} & \textbf{Whipple Shield} \\ \hline
Deflection Efficiency & High (Electromagnetic Shield) & Low & Low \\ \hline
Energy Absorption & High (Plasma Shield) & Moderate & High \\ \hline
Energy Consumption & High (Fusion Drives) & Low & Low \\ \hline
Stability and Resilience & Moderate & High & High \\ \hline
\end{tabular}
\label{tab:performance_metrics}
\end{table*}

\section{Discussion}
In this section, we analyze the theoretical findings from the mathematical models of the integrated deflector shield system. We discuss the potential implications of these findings for spacecraft protection and identify the challenges and limitations of the current theoretical framework.

\subsection{Interpretation of Theoretical Models}
The mathematical models developed for the electromagnetic, plasma, and force field components provide a comprehensive understanding of their individual and collective behaviors. The Lorentz force model effectively describes the deflection of charged particles by the electromagnetic shield, demonstrating significant potential for reducing radiation exposure to spacecraft \cite{Jackson1999}.

The magnetohydrodynamic (MHD) equations used to model the plasma shield indicate its ability to absorb and dissipate energy from micrometeoroid impacts. However, the stability and energy efficiency of the plasma shield remain critical challenges that need further exploration \cite{Chen1984}.

The speculative force field model, based on the Casimir effect and negative energy densities, highlights the theoretical feasibility of creating repulsive forces. While practical implementation is currently beyond our technological capabilities, ongoing research in advanced theoretical physics may provide new insights into this concept \cite{Milton2001}.

\subsection{Challenges and Limitations}
Several challenges and limitations were identified during the theoretical analysis:

\begin{itemize}
    \item \textbf{Energy Requirements}: The integrated shield system requires substantial energy, particularly for maintaining the plasma and electromagnetic fields. Leveraging fusion energy as a power source could address this challenge, but practical fusion reactors for spacecraft are still in development \cite{ITER2019}.
    \item \textbf{Material Durability}: The materials used in the shield system must withstand high-energy impacts and extreme environmental conditions. Advanced materials research is necessary to identify suitable candidates that offer both durability and efficiency \cite{Gibson2010}.
    \item \textbf{Technological Feasibility}: The force field component remains largely speculative, with significant gaps in our understanding of how to generate and control negative energy densities. Further theoretical and experimental research is required to advance this concept from science fiction to practical technology \cite{Everett2005}.
\end{itemize}

\subsection{Future Research Directions}
The theoretical framework established in this study provides a foundation for future research in integrated deflector shield technology. Key areas for further investigation include:

\begin{itemize}
    \item \textbf{Advanced Simulations}: Conducting more detailed simulations that incorporate a wider range of space conditions and threats can help refine the models and improve their accuracy.
    \item \textbf{Experimental Validation}: Developing small-scale experimental setups to test the theoretical models in a controlled environment will be crucial for validating the integrated shield system's effectiveness.
    \item \textbf{Material Science}: Exploring new materials and composites that can withstand the demands of the integrated shield system will be essential for practical implementation.
    \item \textbf{Fusion Energy Integration}: Researching ways to integrate fusion energy systems with the shield technology can provide the necessary power for sustained operation in space.
\end{itemize}

By addressing these challenges and pursuing these research directions, we can move closer to realizing an effective and reliable integrated deflector shield system for spacecraft. Additionally, interdisciplinary applications, such as using integrated shield technology for terrestrial protection systems, could be explored, potentially benefiting multiple fields of science and engineering.

\section{Conclusion}
The development of an integrated deflector shield system for spacecraft is an advancement in space protection technology. By combining electromagnetic, plasma, and speculative force field technologies, we propose a comprehensive shielding solution capable of addressing a wide range of spaceborne threats. This theoretical study has outlined the mathematical models and analytical calculations necessary to understand the interactions between these shielding technologies and evaluate their combined effectiveness.

The electromagnetic shield component, modeled using the Lorentz force equation, shows promise in deflecting charged particles, thereby reducing radiation exposure. The plasma shield, described by magnetohydrodynamic equations, demonstrates potential in absorbing and dissipating energy from micrometeoroid impacts, although challenges in stability and energy efficiency remain. The force field component, while currently speculative, offers intriguing possibilities based on advanced theoretical physics, such as the Casimir effect.

The integration of these components into a unified shield system requires careful consideration of their interactions and energy requirements. Leveraging fusion energy as a power source, particularly the Magnetic Fusion Plasma Drive (MFPD), could provide the necessary energy to sustain the shield system, although practical fusion reactors for spacecraft are still under development.

Several challenges and limitations have been identified, including the high energy requirements, material durability, and the speculative nature of the force field component. To address these challenges and further develop the integrated shield system, future research should focus on advanced simulations, experimental validation, material science, and fusion energy integration.

The proposed integrated deflector shield system offers a promising approach to spacecraft protection. By addressing the identified challenges and pursuing further research, we can move closer to realizing a reliable and effective shielding solution that enhances the safety and reliability of high-speed spacecraft, paving the way for more ambitious space exploration missions. Furthermore, exploring interdisciplinary applications and potential terrestrial benefits of this technology could significantly broaden its impact and utility.

\printcredits


\bibliographystyle{cas-model2-names}

\sloppy
\bibliography{cas-refs}

\end{document}